\documentclass[aps,prl,reprint,groupedaddress]{revtex4-1}
\usepackage{graphicx}
\usepackage{amssymb,amssymb}
\usepackage{lineno}

\usepackage{subfigure}

\newlength{\figwidth}
\RequirePackage[colorlinks,breaklinks,pdftitle={CERN-PH-EP-2010-033},pdfauthor={The ATLAS Collaboration}]{hyperref}
\hypersetup{linkcolor=blue,citecolor=blue,filecolor=black,urlcolor=blue}

\begin{document}

% CERN-PH: CERN-PH-EP-2010-033   for the contact interactions

\begin{flushleft}
{\bf CERN-PH-EP-2010-033}
\end{flushleft}

\begin{flushright}
%{\bf 25 September 2010}
\end{flushright}

%---------------------------- TEX DEFS ---------------------------------

\def\integLumi{3.1~pb$^{-1}$}
\def\integLumiUncertainty{$\pm 11$\%}
\def\JERuncertainty{14} % NB: must manually tweak app:Systematics; no \pm or percentage indicated!
\def\MjjCut{200} % Note: no units attached!
\def\MjjLowerExclusion{400} % Note: no units attached!
\def\MjjUpperFitRange{1900} % Note: no units attached!
\def\MjjLowerLimitMRST{1265} % Note: no units attached!
\def\MjjLowerLimitMRSTEXPECTED{1100} % Note: no units attached!
\def\MjjLowerLimitStatOnlyMRST{1295} % Note: no units attached!
\def\MjjLowerLimitStatOnlyMRSTEXPECTED{1140} % Note: no units attached!
\def\MjjLowerLimitCTEQsixLone{1205} % Note: no units attached!
\def\MjjLowerLimitCTEQsixLoneEXPECTED{1020} % Note: no units attached!
\def\MjjLowerLimitStatOnlyCTEQsixLone{1240} % Note: no units attached!
\def\MjjLowerLimitStatOnlyCTEQsixLoneEXPECTED{xxxx} % Note: no units attached!
\def\MjjLowerLimitCTEQsixsix{1245} % Note: no units attached!
\def\MjjLowerLimitCTEQsixsixEXPECTED{1060} % Note: no units attached!
\def\MjjLowerLimitStatOnlyCTEQsixsix{1280} % Note: no units attached!
\def\MjjLowerLimitStatOnlyCTEQsixsixEXPECTED{xxxx} % Note: no units attached!
\def\NpseudoExperiments{$10^3$}
\def\NpseudoExperimentsQCDfit{$10^4$}
\def\triggerThreshold{15} % Note: no units attached!
\def\pTcut{80} % Note: no units attached!
\def\topbottomLabel{~}
\def\leftrightLabel{~}
\def\Pythia{{\sc Pythia}}
\def\GeVcc{${\rm GeV}$}
\def\TeVcc{${\rm TeV}$}
\def\GeVc{${\rm GeV}$}

\def\pt{p$_t$}

%-------------------------- Start of Document -----------------------------

%***\begin{document}

%\linenumbers

%\preprint{}

\title{Search for Quark Contact Interactions in Dijet Angular Distributions in 
  $pp$ Collisions at $\sqrt{s}=7$~TeV Measured with the 
  ATLAS Detector}


\author{The ATLAS Collaboration}
%\thanks{}
\affiliation{}


\date{\today}

% Doesn't work. Needs to be in a .cls file.
%\renewcommand{\section}{\@startsection {section}{1}{Opt}%
%     {10pt}%
%     {-3.5ex plus -1ex minus -.2ex}%
%     {2.3ex plus .2ex}%
%     {\normalfont\normalsize\bfseries}}

%----------------------------- Abstract -----------------------------

\begin{abstract}
Dijet angular distributions from the first LHC $pp$ collisions at center-of-mass
energy $\sqrt{s}=7$~TeV have been measured with the ATLAS detector.
The dataset used for this analysis represents an integrated luminosity of \integLumi. 
Dijet $\chi$ distributions and centrality ratios have been measured up to
dijet masses of 2.8 TeV, and found to be in good agreement with 
Standard Model predictions.  Analysis of the $\chi$ distributions excludes
quark contact interactions with a compositeness scale $\Lambda$ below 3.4 TeV, 
%3.43 TeV
at $95\%$ confidence level, significantly exceeding previous limits.
\end{abstract}

%\keywords{}

\maketitle

%------------------------------ Body of Paper -------------------------

%%%\section{}
%%%\subsection{}
%%%\subsubsection{}

%\section{Introduction}

\bigskip
\noindent {\bf 1. Introduction}
\smallskip

At hadron colliders, most events with large transverse momentum ($p_T$) 
transfer occur when a constituent parton from one of the incoming hadrons
scatters from a parton in the other.  At high $p_T$, these `2~$\to$~2'
scattering processes are well described within the Standard Model by perturbative
Quantum Chromodynamics (QCD), the quantum field theory of strong interactions.  
As each high-momentum parton emerges from the collision, the subsequent parton
shower and hadronization create a collimated jet of particles aligned with
the direction of the original parton. In most of these collisions, two high-$p_T$
jets emerge from the interaction. These `dijet' events are particularly
useful for measuring quantities associated with the initial
interaction, such as the polar scattering angle in the two-parton
center-of-mass (CM) frame, $\theta^{*}$, and the dijet invariant mass, $m_{jj}$.
Precise tests of QCD may be carried out by comparing the theoretical predictions
to the experimental distributions. If discrepancies between data and QCD are
found to be well beyond experimental and theoretical uncertainties,
this would indicate that the QCD description needs improvement,
or that a new process, not included in the Standard Model, has appeared. 

This analysis focuses on dijet angular distributions, which have been shown
by previous experiments \cite{Arnison1986244,Bagnaia1984283,CDF_latest,collaboration-2009-103}
to be sensitive measures for testing the predictions of QCD and searching for new
processes. Dijet angular distributions are well suited to the
analysis of early LHC data, since they are little affected by the main
systematic uncertainties associated with the jet energy
scale (JES) and the luminosity. QCD calculations predict that high-$p_T$ dijet
production is dominated by $t$-channel gluon exchange, leading to angular distributions
that are peaked at $|\mathrm{cos}\theta^{*}|$ close to 1.
By contrast, models of new processes characteristically predict angular
distributions that would be more isotropic than those of QCD.

This Letter reports on the first search with the ATLAS detector for quark contact interactions
leading to modifications of dijet angular distributions in proton-proton ($pp$) collisions at
a center-of-mass energy of $\sqrt{s} = 7$~TeV at the LHC.  The data sample represents
an integrated luminosity of \integLumi, recorded in periods of stable collisions, through
August 2010. The two distributions under study - dijet $\chi$ distributions, 
and dijet centrality ratios - have been used repeatedly as benchmark measures,
and will be described in detail below.

The highest exclusion limits on quark contact interactions set by any
previous experiment \cite{collaboration-2009-103}, for several
statistical analyses, ranged from 2.8 to 3.1 TeV at 95\% CL for
the compositeness scale $\Lambda$.  

%\section{Kinematics and Angular Distributions}

\bigskip
\noindent {\bf 2. Kinematics and Angular Distributions}
\smallskip

The $\theta^{*}$ distribution for $2\to2$ parton scattering is predicted by
QCD in the partonic CM frame of reference.
Event by event, the momentum fraction (Bjorken $x$)
of one incoming parton differs from that of the other, causing the partonic
reference frame to be boosted relative to the detector frame by an
amount which can be determined from the dijet kinematics.
A natural variable for analysis of parton-parton interactions is therefore the rapidity, 
$y = \frac{1}{2}\mathrm{ln}(\frac{E + p_z}{E - p_z})$,
where $E$ is the energy and $p_z$, the $z$-component of momentum, of the given particle.
The variable $y$ transforms under Lorentz boosts along the $z$-direction as
$y \to y - y_B = y - \mathrm{tanh}^{-1}(\beta_B)$,
where $\beta_B$ is the velocity of the boosted frame,
and $y_B$ is its rapidity boost. 

The ATLAS coordinate system is a right-handed Cartesian
system with the $x$-axis pointing to the center of the LHC ring, the $z$-axis
following the counter-clockwise beam direction, and the $y$-axis going upwards.
The polar angle $\theta$ is referred to the $z$-axis, and $\phi$ is the azimuthal
angle about the $z$-axis.

Rapidity differences are boost invariant,
so that under Lorentz boosts jets retain their shapes in ($y$,$\phi$) coordinates.
The pseudorapidity, $\eta = -\mathrm{ln}(\mathrm{tan}(\frac{\theta}{2}))$, 
approaches rapidity in the massless limit and can be used as an approximation 
to rapidity. The variables $\eta$ and $\phi$ are employed in the reconstruction
of jets.

The variable $\chi$, used in the first angular distributions considered in 
this study, is derived from the rapidities of the two jets defining the 
dijet topology ($y_1$ and $y_2$).
For a given scattering angle $\theta^*$, the corresponding rapidity in the CM frame
(in the massless particle limit) is
$y^* = \frac{1}{2}\mathrm{ln}(\frac{1 + |\mathrm{cos}\theta^*|}{1 - |\mathrm{cos}\theta^*|})$.
The variables $y^*$ and $y_B$ can be found from the rapidities of the two jets using
$y^* = \frac{1}{2}(y_1 - y_2)$ and $y_B = \frac{1}{2}(y_1 + y_2)$.
Then $y^*$ may be used to determine the partonic CM angle $\theta^*$.  Additionally,
$y^*$ is the basis for the definition of $\chi$:
$\chi =\text{exp}(|y_1-y_2|)=\text{exp}(2|y^*|)$.

The utility of the $\chi$ variable becomes apparent when making comparisons
of angular distributions predicted for new processes to those of QCD.
In QCD, gluon (massless, spin-1) exchange diagrams have approximately the same angular 
dependence as Rutherford scattering: 
$\mathrm{d}N / \mathrm{d cos}\theta^* \propto 1/\mathrm{sin}^4(\theta^*/2)$.
Evaluation of $\mathrm{d}N / \mathrm{d}\chi$ shows that this
distribution is constant in $\chi$. 
By contrast, the angular distributions characteristic of new processes
are more isotropic, leading to additional dijet events at low $\chi$.
In QCD, subdominant diagrams also cause $\chi$ distributions to rise 
slightly at low $\chi$.

The other important kinematic variable derivable from jet observables is the
dijet invariant mass, $m_{jj}$, which is also the CM energy of the
partonic system. In reconstruction, $m_{jj}$ is found from the two jet four-vectors:
$m_{jj} \equiv \sqrt{(E^{ j_1} + E^{j_2})^2 - (\vec{p}^{j_1} + \vec{p}^{j_2})^2}$,
where $E$ and $\vec{p}$ are the energy and momentum of the jets.  Both 
distributions used in this Letter are binned in this variable.

The second angular distribution considered is the dijet centrality ratio, $R_C$. For this
analysis, the detector is divided into two pseudorapidity regions: central and non-central.
$R_C$ is defined as the ratio of the number of events in which the two highest
$p_T$ jets both fall into the central region to the number of events in which the
two highest $p_T$ jets both fall into the non-central region. 
For the current study, the central region is defined as
$|\eta_{1,2}| < 0.7$, and the non-central region as $0.7 < |\eta_{1,2}| < 1.3$. 
Since new processes are expected to produce more central activity than QCD, 
their signal would appear as an increase in $R_C$ above some $m_{jj}$ threshold,
with the increase being directly related to the cross section of the new signal.

$R_C$ distributions are complementary to $\chi$ distributions by being sensitive
to different regions of phase space. $\chi$ distributions are fine measures of
$\theta^*$ and coarse measures of $m_{jj}$, while the opposite is true for $R_C$ distributions
as they can be binned more finely in $m_{jj}$ for the given amount of data.
This gives $R_C$ distributions greater discrimination in determining mass scales
associated with hypothetical signals.  Ideally, when a signal is present,
the two distributions together can be used to narrow the list of viable hypotheses
and to establish the associated scale parameters.

It should be noted that measured $\chi$ and $R_C$ distributions are compared to theoretical
predictions after jet reconstruction in the detector, which smears the
parton-level distributions.

%\section{The ATLAS Detector}

\newpage
\bigskip
\noindent {\bf 3. The ATLAS Detector}
\smallskip

The ATLAS detector~\cite{DetectorPaper} covers almost the whole solid angle around 
the collision point with layers of tracking detectors, calorimeters, and muon chambers.
Jet measurements depend most strongly on the calorimeter system.  The ATLAS calorimeter is 
segmented in intervals of pseudorapidity and $\phi$ to exploit the property that jet shapes
are nearly boost invariant in ($\eta$,$\phi$) coordinates.

Liquid argon (LAr) technology is used in the electromagnetic sampling calorimeters,
with excellent energy and position resolution, to cover the pseudorapidity range $|\eta| < 3.2$.
The hadronic calorimetry in the range $|\eta| < 1.7$ is provided by a sampling calorimeter 
made of steel and scintillating tiles. 
In the end-caps ($1.5 < |\eta| < 3.2$), LAr technology is also used for the hadronic calorimeters,
matching the outer $|\eta|$ limits of the electromagnetic calorimeters. 
To complete the $\eta$ coverage, the LAr forward calorimeters provide both
electromagnetic and hadronic energy measurements, extending the coverage to $|\eta| = 4.9$.
The ($\eta$,$\phi$) segmentation of the calorimeter is sufficiently fine to assure
that angular resolution uncertainties in dijet analyses are negligible.
In the data taking period considered approximately 187,000 calorimeter cells (98\% of the total)
were active for event reconstruction.

ATLAS has a three-level trigger system, with the first level (L1) being custom built hardware.
The two higher level triggers (HLT) are realized in software.  The HLT was not set to reject events
accepted by the L1 single-jet triggers chosen for this analysis.

%\section{Reconstruction and Event Selection}

%for registration of headings
%\newpage

\bigskip
\noindent {\bf 4. Event Selection and Reconstruction}
\smallskip

In the current \integLumi data sample,
specific L1 jet trigger selections have been exploited for optimal
analysis of the angular observables.
For both observables, bins of $m_{jj}$ are associated with distinct 
L1 jet trigger requirements selected to provide maximal statistics while being
fully efficient, as will be detailed for $\chi$ in Section 7.

Jets have been reconstructed using the infrared-safe anti-$k_t$ jet clustering 
algorithm~\cite{antikT} with the radius parameter $R = 0.6$. The inputs to this
algorithm are clusters of calorimeter cells seeded by energy depositions significantly
above the measured noise.  Jet four-vectors are constructed by the vectorial addition of
cell clusters, treating each cluster as an ($E$, $\vec{p}$) four-vector with zero mass. 
The jet four-vectors are then corrected, as a function of $\eta$ and $p_T$, for the effects
of hadronic shower response and detector material distributions by using a calibration scheme
based on Monte Carlo (MC) studies including full detector simulation, and validated 
with extensive test-beam studies~\cite{Adragna:2010zz} and collision data \cite{JES}.

In order to suppress cosmic-ray and beam-related backgrounds, 
events are required to contain at least one primary collision vertex with a position
of $\left|z\right| < 30$~cm and reconstructed from at least five charged-particle tracks.
Events with at least two jets are retained if the highest $p_T$\ jet
(the `leading' jet) satisfies $p_T^{j_1} > 60$~GeV and the next-to-leading
jet satisfies $p_T^{j_2} > 30$~GeV. 
The asymmetric thresholds are required so as to avoid suppression of events where
a third jet has been radiated, while the 30~GeV threshold ensures that reconstruction
is fully efficient for both leading jets.  Those events containing a
poorly measured jet \cite{JetEtMiss_Cleaning_Note} with $p_T > 15$~GeV are vetoed to avoid
cases where such a jet would cause incorrect identification of the two leading jets.
This criterion results in a rejection rate of 0.5\% in the current data sample.
The two leading jets are required to satisfy quality criteria, such as being
associated with in-time energy deposits in the calorimeter.
To be considered as one of the two leading jets, a jet is required to be found within
the pseudorapidity region $|\eta| < 2.8$, where the jet energy scale is known to
highest precision.

%\section{Simulation}

\bigskip
\noindent {\bf 5. Monte Carlo Simulations}
\smallskip

The Monte Carlo simulation used for the analysis presented in this Letter
has the following components.  The MC
samples have been produced with the PYTHIA 6.4.21 event generator \cite{pythia} and the
ATLAS MC09 parameter tune 
\cite {ATLAS_MC09}, using the modified leading-order MRST2007 \cite{Sherstnev:2007nd}
parton distribution functions (PDF).
The generated events are passed through the detailed simulation of the ATLAS detector
\cite{ATLSIM}, which uses GEANT4 \cite{Agostinelli:2002hh} for simulation of particle 
transport, interactions, and decays.
This yields QCD MC samples that have been smeared by detector effects for
comparison with collision data.
These simulated events are then subjected to the same reconstruction process
as the data to produce dijet angular distributions.

As the next step, bin-by-bin correction factors (K-factors) have been applied
to the angular distributions derived from MC events to account
for next-to-leading order (NLO) contributions. The K-factors are defined as
the ratio $NLO_{ME}$/$PYT_{SHOW}$.  The $NLO_{ME}$ sample is produced 
using matrix elements in NLOJET++ \cite{catani-1998-510} and the NLO
PDF from CTEQ6.6 \cite{Nadolsky:2008zw}. The $PYT_{SHOW}$ sample
is produced with PYTHIA restricted to leading-order (LO) matrix elements
and parton showering using the modified leading order MRST2007 PDF.
Multiplying the full PYTHIA predictions of
angular distributions with the bin-wise K-factors, obtained as described
above, results in a reshaped spectrum which includes corrections originating
from NLO matrix elements, while retaining the non-perturbative effects
simulated in PYTHIA.  Over the full range of $\chi$, the K-factors change 
the normalized distributions by up to 6\%, with little variability
from one mass bin to the other. In the case of $R_C$, the K-factors change
the distribution by less than 1\%.

The QCD predictions used for comparison to data in this Letter are
the end product of the two-step procedure described above.

%MOD-SEPT22
Other ATLAS jet studies \cite{jetProdICHEP} have shown that the use of 
different event generators and different sets of parameters for
non-perturbative effects (e.g. hadronization) has a negligible effect on
the studied observables in the phase space being considered. 
For the high-$p_T$  dijet shape observables studied here,
$\chi$  and $R_C$, differences between PDF sets were found to be consistently
smaller than the uncertainty associated with the CTEQ6.6 PDF set, and 
are not taken into account.

%\section{Quark Contact Term Interaction}
%\newpage

\bigskip
\noindent {\bf 6. Quark Contact Interaction Term}
\smallskip

The benchmark beyond-the-Standard-Model process considered in this Letter is a quark
contact interaction, which may be used to model the onset of kinematic properties that
would characterize quark compositeness:
the hypothesis that quarks are composed of more fundamental particles.
The model Lagrangian for this benchmark process is a four-fermion contact interaction 
\cite{Eichten:1984eu,Eichten:1995akc,Chiappetta1991},
the analog of the Fermi four-fermion interaction used to describe effects
of the weak interaction.
The effects of the contact interaction would be
expected to appear below or near a characteristic energy scale $\Lambda$.  
If $\Lambda$ is much larger than the partonic CM energy, these interactions are suppressed
by inverse powers of $\Lambda$ and the quarks would appear to be point-like. The dominant effect
would then come from the lowest dimensional four-fermion interactions (contact terms). 

While a number of contact terms are possible, the Lagrangian in standard use since
1984 \cite{Eichten:1984eu} is the single (isoscalar) term: 
$\mathcal{L}_{qqqq}(\Lambda)=\frac{{\xi}g^2}{2\Lambda^2_q}
\bar{\Psi}^L_q\gamma^{\mu}\Psi^L_q\bar{\Psi}^L_q\gamma_{\mu}\Psi^L_q$,
where $g^2/4\pi=1$ and the quark fields
$\Psi^L_q$ are left-handed. The full Lagrangian used for hypothesis testing is
then the sum of $\mathcal{L}_{qqqq}(\Lambda)$ and the QCD Lagrangian.  
The relative phase of these terms is controlled by the interference parameter,
$\xi$, which is set for destructive interference ($\xi = +1$) in the current analysis.
Previous analyses \cite{collaboration-2009-103} showed that the choice of 
constructive ($\xi = -1$) or destructive ($\xi = +1$) interference changed exclusion 
limits by $\sim 1\%$.

MC samples are calculated in PYTHIA 6.4.21 using this Lagrangian, with each
sample corresponding to a distinct value of $\Lambda$.  Angular distributions of
these samples are processed in the same fashion as QCD distributions, including
the application of bin-wise K-factors.  
%Lacking a theoretical prescription for
%applying NLO corrections to contact interactions, this is the only sensible
%method.  This method is expected to be stable because for small compositeness signals
%the angular distributions approach the QCD predictions.

Notably, in addition to quark compositeness, this same Lagrangian could be applied to a 
number of other beyond-the-Standard-Model theories (albeit with different coupling constants),
so that it serves as a template for models of new processes with similar scattering
distributions.

%\section{Kinematic Criteria and Binning of Angular Distributions}

\bigskip
\noindent {\bf 7. Kinematic Criteria for Angular Distributions}
\smallskip

%MOD-SEPT22
The $\chi$ distributions described here are normalized to unit area,
$(1/N_{\mathrm{ev}}) \mathrm{d}N_{\mathrm{ev}}/ \mathrm{d} \chi$
where $N_{\mathrm{ev}}$ is the number of observed events,
to reduce the effects of uncertainties associated with absolute
normalization.

Detector resolution effects smear
the $\chi$ distributions, causing events to migrate between neighboring bins.
This effect is reduced by configuring the $\chi$ bins to match the natural 
segmentation of the calorimeter, by making them intervals of constant $\Delta\eta$,
approximating $\Delta y$.  This is achieved by placing the $\chi$ bin
boundaries at positions $\chi_n = \mathrm{e}^{(0.3 \times n)}$, where $n$ is the index for
the lower $\chi$ boundary of the $n^{\mathrm{th}}$ bin, starting with $n = 0$. In doing this, not only
is the migration reduced, it is also equalized across the span of $\chi$.

The $\chi$ distributions are divided, in turn, into intervals of dijet invariant 
mass, $m_{jj}$. For massless partons, the following approximate form
shows the dependence of $m_{jj}$ on $p_T$ and
$\chi$: $m_{jj} = \sqrt{p_{T1} \, p_{T2} }\cdot\sqrt{ \chi+1/\chi -2 \cos(\Delta\phi) }$.
Since $m_{jj}$ is the CM energy of the partonic system, new processes
would be expected to appear in the high mass bins.  Several $m_{jj}$ 
intervals are analyzed to exploit the fact that the $\chi$ distributions 
in low mass bins would be similar to the QCD prediction, while these distributions
would be modified by new physics processes acting in high $m_{jj}$ bins.
The sensitivity to these processes depends strongly on their cross sections
relative to QCD and on the number of events in the highest mass bin.

For $\chi$ distributions, events are rejected if $|y_B| > 0.75$ or $|y^*| > 1.7$.
The combined criteria limit the rapidity range of both jets to $|y_{1,2}| < 2.45$.  
The $|y^*|$ criterion determines the maximum $\chi$, which is 30 for this analysis.
Taken together, these two criteria define a region within the space
of accessible $y_1$ and $y_2$ where the acceptance is
almost flat in $\chi$.  This ensures that the expected shapes of the distributions
are not significantly changed by the acceptance.  
Also, in low mass bins, the $y_B$ criterion emphasizes the contribution from
the matrix elements and reduces the influence of the effects of PDF convolution.
In the highest $m_{jj}$ bin, used for limit
setting, the $|y_B|$ criterion reduces the sample by 16\%.
These kinematic cuts have been optimized through full MC simulation to assure high
acceptance in all dijet mass bins.  

Since event migration also occurs between bins of $m_{jj}$, studies of
fully simulated jets are used to ensure that migration is small. 
This criterion, along with the requirement of a sufficient number of events, lead
to $m_{jj}$ bin boundaries of 340, 520, 800, and 1200~GeV, with no upper
bound on the highest bin. As noted earlier, single-jet triggers are carefully
selected for each bin to be 100\% efficient.  Prescaling of triggers leads to a
different effective integrated luminosity in each mass bin, with the
corresponding numbers being 0.12, 0.56, 2.0, and 3.1~pb$^{-1}$  in the
current data sample for the bins listed above.

Like the $\chi$ distributions, the $R_C$ distribution has reduced sensitivity
to the absolute JES.  However, relative differences in jet response in $\eta$
could have a significant impact on the sensitivity.
Hence, for these early studies, the $\eta$ range is restricted to the more
central regions of the calorimeter
where the JES is uniform to within 1\% as determined from cross-calibration studies.
The $R_C$ region has been chosen to end at a maximum of $|\eta|$ = 1.3, just before the
transition region between the central and end-cap calorimeters.
%The $R_C$ distributions are finely binned in $m_{jj}$, starting from 250~GeV,
%to be able to resolve potential features.

%\section{Systematic Uncertainties}

\bigskip
\noindent {\bf 8. Convolution of Systematic Uncertainties}
\smallskip

As mentioned before, the angular distributions have a 
reduced sensitivity to the JES uncertainties compared to other dijet 
measurements. Nevertheless, the JES still represents the dominant 
uncertainty for this analysis.
The ATLAS JES has been determined by extensive studies \cite{JES2}, and
its uncertainty has been tabulated in the variables $\eta$, $p_T$, and $N_V$,
the number of vertices in the event. The average $N_V$ over the full current data sample
is 1.7.  Typical values of the JES uncertainty in the considered
phase space are between 5\% and 7\%. The resulting bin-wise
uncertainties are up to 9\% for the $\chi$ observable, and
up to 7\% for the $R_C$ observable.

The dominant sources of theoretical uncertainty are NLO QCD
renormalization and factorization scales, and the PDF uncertainties.
The corresponding bin-wise uncertainties for normalized $\chi$ distributions are
typically up to 3\% for the combined NLO QCD scales and 1\% for the PDF error. 

Convolution of these experimental and theoretical uncertainties is done
for all angular distributions through Monte Carlo pseudo-experiments (PE's). 
For all events in the MC sample 1000 PE's are performed, three random numbers being drawn from 
a Gaussian distribution for each PE.  The first is applied
to the absolute JES, obtained from the tabulation described above and assumed to be
fully correlated across $\eta$. The second number is applied to the 
relative JES, extracted from the same tabulation, which depends only on $\eta$ and restores
the decorrelation due to $\eta$ dependence of the energy scale.  The third number is applied
to the PDF uncertainty, provided by the CTEQ6.6 PDF error sets.  In a fourth and final step, the
uncertainty due to the NLO QCD renormalization ($\mu_R$) and factorization ($\mu_F$) scales is
found by letting $\mu_R$ and $\mu_F$ vary independently between 0.5, 1 and 2 times the
average transverse momentum of the two leading jets, resulting in
nine samples drawn from a uniform distribution. In a given PE, the data
dijet selection criteria described previously are applied.

%\section{Comparison of Data with Theory}

\begin{figure}
%\begin{center}
%\subfigure{\includegraphics[width=0.5\figwidth]{figures/chi_data_vs_mc.eps}}
\subfigure{\includegraphics[width=0.5\figwidth]{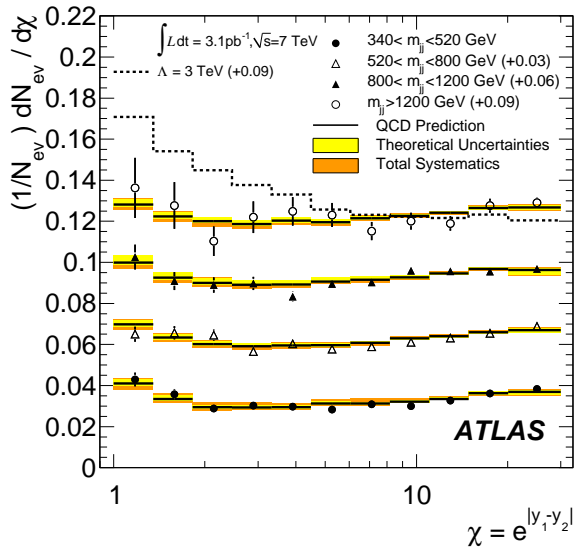}}
%\end{center}
\caption{The normalized $\chi$ distributions for $340 < m_{jj} < 520$~GeV,
         $520 < m_{jj} < 800$~GeV, $800 < m_{jj} < 1200$~GeV, and $m_{jj} > 1200$~GeV,
         with plotting offsets shown in parentheses.
         Shown are the QCD predictions with
         systematic uncertainties (bands), and data points
         with statistical uncertainties.
         The prediction for QCD with an added quark contact term with $\Lambda$ = 3.0~TeV
         is shown for the highest mass bin $m_{jj} > 1200$~GeV.}
%\label{fig:ExclusionWithStatSystematics}
\end{figure}

\bigskip

%\newpage
\noindent {\bf 9. Comparison of Data to Theory}
\smallskip

In Fig. 1 the measured dijet $\chi$ distributions are
compared to the QCD predictions, along with 1 $\sigma$ systematic error bands determined
from the PE's, and statistical errors on the data.  Figure 2 shows the measured dijet-centrality
distribution and QCD prediction.  The statistical uncertainties are obtained using
Poisson probabilities.  In the highest mass bins, the numerator and
denominator of the ratio typically contain 1 or 2 events each.

To evaluate the agreement between data and QCD in Figs. 1 and 2,
chi-square goodness-of-fit tests were performed on
each angular distribution under the assumption that the bin-to-bin
correlations are negligible. 
For the $\chi$ distributions shown in Fig. 1, the chi-square per degree of freedom
for each dijet mass bin is (from lowest to highest) 0.68, 0.83, 0.72, and 0.81,
indicating good agreement with the QCD prediction.  

Similarly, in Fig. 2 the dijet
$R_C$ comparison has a chi-square per degree of freedom equal to 0.61, 
also indicating good agreement with the QCD prediction.

The best fit of the $R_C$ distribution in Fig. 2 is obtained for a
compositeness scale of 2.9~TeV. This is not statistically significant,
as the QCD prediction lies within the shortest 68\% confidence
interval in $1/\Lambda^4$.

%\section{Determining Exclusion Limits}

\bigskip
\noindent {\bf 10. Determination of Exclusion Limits}
\smallskip

\begin{figure}
%\begin{center}
\subfigure{\includegraphics[width=0.5\figwidth]{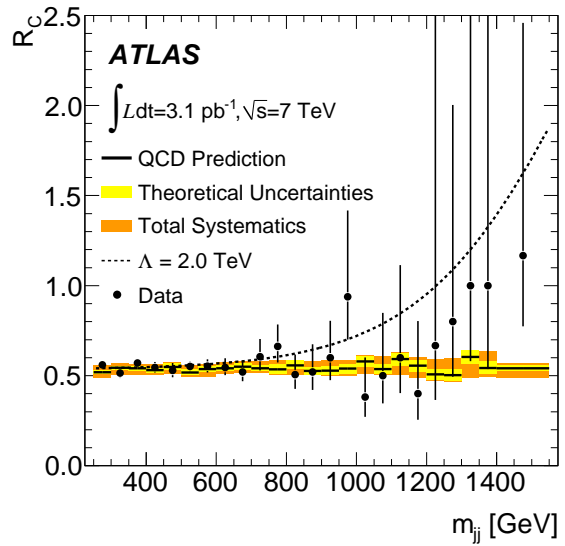}}
%\end{center}
\caption{Dijet centrality ratio, $R_C$, as a function of 
         $m_{jj}$, with all events above a mass of $1400$~GeV 
         plotted in the last bin.  Shown are the QCD 
         prediction with systematic uncertainties (bands), and data points
         with statistical uncertainties.
         The prediction for QCD with an added quark contact term 
         with $\Lambda$ = 2.0~TeV is also shown.}
%\label{fig:ExclusionWithStatSystematics}
\end{figure}

Since no signal from new physics processes is apparent in these distributions,
limits have been obtained on the compositeness scale $\Lambda$ of
quark contact interactions, based on analyses of the $\chi$ distributions.
The contact term hypothesis is
tested in the highest dijet mass bin in Fig. 1, which begins at $m_{jj} = 1200$~GeV.
For the $\chi$ distribution in this mass bin, the parameter $F_{\chi}$ is
defined as the ratio of the number of events
in the first four $\chi$ bins to the number in all $\chi$ bins.
The upper boundary of the fourth bin is at $\chi = 3.32$.  This choice of
the bin boundary has been determined through a MC study that varies the
number of bins in the numerator, as well as the dijet mass bin, and 
determines the setting that maximizes the sensitivity to quark contact
interactions, given the current integrated luminosity.

A frequentist analysis is employed as follows. Predictions of $F_{\chi}$ are
obtained for a range of $\Lambda$ by interpolation between
distinct samples generated with different $1/\Lambda^2$ values. 
The QCD sample provides a bound with $\Lambda = \infty$,
and additional samples are generated with $\Lambda$ values of
500, 750, 1000, 1500, and 3000~GeV.  A full set of
PE's is made for each hypothesis to construct one-sided $95\%$
confidence level (CL) intervals for $F_{\chi}$, and the Neyman
construction \cite{Neyman} is then applied to obtain a limit on $\Lambda$.
  
The result is shown in Fig. 3.
The measured value of $F_{\chi}$ is shown by the dashed horizontal line.
The value of $F_{\chi}$ expected from QCD is the solid horizontal line,
and the band around it allows one to obtain the 1 $\sigma$ variation of the expected limit.
The dotted line is the $95\%$ CL contour of the $F_{\chi}$ prediction for quark
contact interactions plus QCD, as a function of $\Lambda$ and including all systematic uncertainties.
This contour decreases as a function of $\Lambda$
since, for a small $\Lambda$ scale, there would be more events at low $\chi$.

The observed limit on $\Lambda$ is 3.4~TeV.  This limit is found from the point
where the $F_{\chi}$ $95\%$ CL contour crosses the measured $F_{\chi}$ value.
All values of $\Lambda$ less than this value are excluded with 95\% confidence.
This corresponds to a distance scale of \mbox{$\sim 6~\cdot$ 10$^{-5}$~fm}, 
from conversion of the limit using $\hbar c$. The expected limit, found from 
the crossing at the QCD prediction, is~3.5~TeV.

%The observed limit on $\Lambda$, found from the point where the $F_{\chi}$ %95\%$ CL
%contour crosses the measured $F_{\chi}$ value, is 3.4~TeV, excluding all values of $\Lambda$
%less than this value with 95\% confidence. 

The impact of systematic uncertainties is as follows.
If all systematic uncertainties
were excluded, the observed limit reported above would increase by 6\% to 3.6~TeV,
mainly due to the JES uncertainty.  Inclusion of NLO scales and PDF
uncertainties does not change the limit measurably, as shape differences arising
from these are well below the statistical uncertainties.

Confirming analyses have been done using a Bayesian approach with Poisson likelihoods 
for $\chi$ bins, calculated using priors flat in $1/\Lambda^2$ or $1/\Lambda^4$.
These have resulted in observed exclusion limits on $\Lambda$ of 3.3~TeV and 3.2~TeV,
respectively, very close to the limit found in the frequentist analysis.

Similarly, an analysis has been performed to establish 95\% CL limits using the
dijet centrality ratio shown in Fig. 2. The likelihood for $R_C$ is constructed as 
a product of likelihoods of inner and outer event counts for all mass bins, which is then
analyzed with a Bayesian approach similar to that of the $\chi$ Bayesian analysis.
Using priors flat in $1/\Lambda^2$ ($1/\Lambda^4$) the observed exclusion limit is
2.0~TeV (also 2.0~TeV), with an expected limit of 2.6~TeV (2.4~TeV), providing
an additional benchmark for comparison with other experiments.
A weaker limit than the one derived from the $\chi$ analysis is expected due to the
lower $\eta$ acceptance associated with the $R_C$ observable.
%and provides
%an additional benchmark for comparison to previous and present analyses 
%concentrating on this observable.

\begin{figure}
%\begin{center}
%\subfigure{\includegraphics[width=0.5\figwidth]{figures/chi_limit_1200.eps}}
\subfigure{\includegraphics[width=0.5\figwidth]{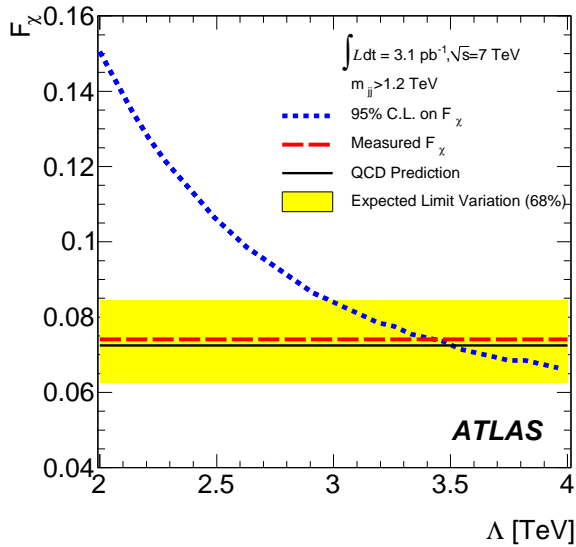}}
%\end{center}
\caption{The dashed horizontal line is the measured $F_{\chi}$ (see text) and
the solid horizontal line is the QCD prediction, with a band to illustrate
a 1 $\sigma$ variation of the expected limit.
The dotted curve is the 95\% CL exclusion contour for $F_{\chi}$ with
quark contact interactions, used to set the exclusion limit on $\Lambda$.
}
%\label{fig:ExclusionWithStatSystematics}
\end{figure}

%\section{Conclusion}

%\bigskip
%\bigskip

%\newpage

\bigskip
\noindent {\bf Conclusion}
\medskip

Dijet angular distributions have been measured by the ATLAS
experiment over a large angular range and spanning dijet masses
up to 2.8~TeV. These distributions are in good agreement
with QCD predictions. Using 3.1~pb$^{-1}$ of data, quark contact interactions
with a scale $\Lambda$ below 3.4~TeV are excluded at the 95\% CL.
The sensitivity of this analysis extends significantly beyond that of 
previously published studies.

%\input{Acknowledgement-20apr10-collisionsPKS.tex}

%\newpage
%\documentclass[11pt,a4paper,dvips]{article}

%\begin{document}

%\section{Acknowledgements}

\bigskip
\bigskip
\bigskip

\raggedright

\noindent {\bf Acknowledgments}
\smallskip

%We deeply thank everybody

We are profoundly grateful to everyone at CERN involved in operating the LHC in such a superb way during this initial high-energy data-taking period. We acknowledge equally warmly all the technical and administrative staff in the collaborating institutions without whom ATLAS could not be operated so efficiently. 

We acknowledge the support of ANPCyT, Argentina; Yerevan Physics Institute, Armenia; ARC and DEST, Australia; Bundesministerium f\"ur Wissenschaft und Forschung, Austria; National Academy of Sciences of Azerbaijan; State Committee on Science \& Technologies of the Republic of Belarus; CNPq and FINEP, Brazil; NSERC, NRC, and CFI, Canada; CERN; CONICYT, Chile; NSFC, China; COLCIENCIAS, Colombia; Ministry of Education, Youth and Sports of the Czech Republic, Ministry of Industry and Trade of the Czech Republic, and Committee for Collaboration of the Czech Republic with CERN; DNRF, DNSRC and the Lundbeck Foundation, Denmark; European Commission, through the ARTEMIS Research Training Network; IN2P3-CNRS and CEA-DSM/IRFU, France; Georgian Academy of Sciences; BMBF, DFG, HGF and MPG, Germany; Ministry of Education and Religion, through the EPEAEK program PYTHAGORAS II and GSRT, Greece; ISF, MINERVA, GIF, DIP, and Benoziyo Center, Israel; INFN, Italy; MEXT, Japan; CNRST, Morocco; FOM and NWO, Netherlands; The Research Council of Norway; Ministry of Science and Higher Education, Poland; GRICES and FCT, Portugal; Ministry of Education and Research, Romania; Ministry of Education and Science of the Russian Federation and State Atomic Energy Corporation ROSATOM; JINR; Ministry of Science, Serbia; Department of International Science and Technology Cooperation, Ministry of Education of the Slovak Republic; Slovenian Research Agency, Ministry of Higher Education, Science and Technology, Slovenia; Ministerio de Educaci\'{o}n y Ciencia, Spain; The Swedish Research Council, The Knut and Alice Wallenberg Foundation, Sweden; State Secretariat for Education and Science, Swiss National Science Foundation, and Cantons of Bern and Geneva, Switzerland; National Science Council, Taiwan; TAEK, Turkey; The STFC, the Royal Society and The Leverhulme Trust, United Kingdom; DOE and NSF, United States of America. 

\bibliography{DijetAngDistPLB}

\begin{thebibliography}{22}%
\makeatletter
\providecommand \@ifxundefined [1]{%
 \@ifx{#1\undefined}
}%
\providecommand \@ifnum [1]{%
 \ifnum #1\expandafter \@firstoftwo
 \else \expandafter \@secondoftwo
 \fi
}%
\providecommand \@ifx [1]{%
 \ifx #1\expandafter \@firstoftwo
 \else \expandafter \@secondoftwo
 \fi
}%
\providecommand \natexlab [1]{#1}%
\providecommand \enquote  [1]{``#1''}%
\providecommand \bibnamefont  [1]{#1}%
\providecommand \bibfnamefont [1]{#1}%
\providecommand \citenamefont [1]{#1}%
\providecommand \href@noop [0]{\@secondoftwo}%
\providecommand \href [0]{\begingroup \@sanitize@url \@href}%
\providecommand \@href[1]{\@@startlink{#1}\@@href}%
\providecommand \@@href[1]{\endgroup#1\@@endlink}%
\providecommand \@sanitize@url [0]{\catcode `\\12\catcode `\$12\catcode
  `\&12\catcode `\#12\catcode `\^12\catcode `\_12\catcode `\%12\relax}%
\providecommand \@@startlink[1]{}%
\providecommand \@@endlink[0]{}%
\providecommand \url  [0]{\begingroup\@sanitize@url \@url }%
\providecommand \@url [1]{\endgroup\@href {#1}{\urlprefix }}%
\providecommand \urlprefix  [0]{URL }%
\providecommand \Eprint [0]{\href }%
\providecommand \doibase [0]{http://dx.doi.org/}%
\providecommand \selectlanguage [0]{\@gobble}%
\providecommand \bibinfo  [0]{\@secondoftwo}%
\providecommand \bibfield  [0]{\@secondoftwo}%
\providecommand \translation [1]{[#1]}%
\providecommand \BibitemOpen [0]{}%
\providecommand \bibitemStop [0]{}%
\providecommand \bibitemNoStop [0]{.\EOS\space}%
\providecommand \EOS [0]{\spacefactor3000\relax}%
\providecommand \BibitemShut  [1]{\csname bibitem#1\endcsname}%
\let\auto@bib@innerbib\@empty
%</preamble>
\bibitem [{\citenamefont {UA1$\;$Collaboration}(1984)}]{Arnison1986244}%
  \BibitemOpen
  \bibfield  {author} {\bibinfo {author} {\bibnamefont
  {UA1$\;$Collaboration}},\ }\href@noop {} {\bibfield  {journal} {\bibinfo
  {journal} {Physics Letters B}\ }\textbf {\bibinfo {volume} {136}},\ \bibinfo
  {pages} {294} (\bibinfo {year} {1984})}\BibitemShut {NoStop}%
\bibitem [{\citenamefont {UA2$\;$Collaboration}(1984)}]{Bagnaia1984283}%
  \BibitemOpen
  \bibfield  {author} {\bibinfo {author} {\bibnamefont
  {UA2$\;$Collaboration}},\ }\href@noop {} {\bibfield  {journal} {\bibinfo
  {journal} {Physics Letters B}\ }\textbf {\bibinfo {volume} {144}},\ \bibinfo
  {pages} {283} (\bibinfo {year} {1984})}\BibitemShut {NoStop}%
\bibitem [{\citenamefont {CDF$\;$Collaboration}(1996)}]{CDF_latest}%
  \BibitemOpen
  \bibfield  {author} {\bibinfo {author} {\bibnamefont
  {CDF$\;$Collaboration}},\ }\href@noop {} {\bibfield  {journal} {\bibinfo
  {journal} {Phys. Rev. Lett.}\ }\textbf {\bibinfo {volume} {77}},\ \bibinfo
  {pages} {5336} (\bibinfo {year} {1996})}\BibitemShut {NoStop}%
\bibitem [{\citenamefont {D0$\;$Collaboration}(2009)}]{collaboration-2009-103}%
  \BibitemOpen
  \bibfield  {author} {\bibinfo {author} {\bibnamefont {D0$\;$Collaboration}},\
  }\href@noop {} {\bibfield  {journal} {\bibinfo  {journal} {Phys. Rev. Lett.}\
  }\textbf {\bibinfo {volume} {103}},\ \bibinfo {pages} {191803} (\bibinfo
  {year} {2009})}\BibitemShut {NoStop}%
\bibitem [{\citenamefont {ATLAS$\;$Collaboration}(2008)}]{DetectorPaper}%
  \BibitemOpen
  \bibfield  {author} {\bibinfo {author} {\bibnamefont
  {ATLAS$\;$Collaboration}},\ }\href {\doibase 10.1088/1748-0221/3/08/S08003}
  {\bibfield  {journal} {\bibinfo  {journal} {JINST}\ }\textbf {\bibinfo
  {volume} {3}},\ \bibinfo {pages} {S08003} (\bibinfo {year}
  {2008})}\BibitemShut {NoStop}%
%%CITATION = JINST,3,S08003;%%
\bibitem [{\citenamefont {Cacciari}\ \emph {et~al.}(2008)\citenamefont
  {Cacciari}, \citenamefont {Salam},\ and\ \citenamefont {Soyez}}]{antikT}%
  \BibitemOpen
  \bibfield  {author} {\bibinfo {author} {\bibfnamefont {M.}~\bibnamefont
  {Cacciari}}, \bibinfo {author} {\bibfnamefont {G.}~\bibnamefont {Salam}}, \
  and\ \bibinfo {author} {\bibfnamefont {G.}~\bibnamefont {Soyez}},\
  }\href@noop {} {\bibfield  {journal} {\bibinfo  {journal} {JHEP}\ }\textbf
  {\bibinfo {volume} {04}},\ \bibinfo {pages} {063} (\bibinfo {year}
  {2008})}\BibitemShut {NoStop}%
\bibitem [{\citenamefont {Adragna}\ \emph {et~al.}(2010)\citenamefont {Adragna}
  \emph {et~al.}}]{Adragna:2010zz}%
  \BibitemOpen
  \bibfield  {author} {\bibinfo {author} {\bibfnamefont {P.}~\bibnamefont
  {Adragna}} \emph {et~al.},\ }\href {\doibase 10.1016/j.nima.2010.01.037}
  {\bibfield  {journal} {\bibinfo  {journal} {Nucl. Instrum. Meth.}\ }\textbf
  {\bibinfo {volume} {A615}},\ \bibinfo {pages} {158} (\bibinfo {year}
  {2010})}\BibitemShut {NoStop}%
%%CITATION = NUIMA,A615,158;%%
\bibitem [{\citenamefont {ATLAS$\;$Collaboration}(2010{\natexlab{a}})}]{JES}%
  \BibitemOpen
  \bibfield  {author} {\bibinfo {author} {\bibnamefont
  {ATLAS$\;$Collaboration}},\ }\href
  {https://atlas.web.cern.ch/Atlas/GROUPS/PHYSICS/RESULTS/summer2010.html}
  {\bibfield  {journal} {\bibinfo  {journal} {ATLAS-CONF-2010-052}\ } (\bibinfo
  {year} {2010}{\natexlab{a}})}\BibitemShut {NoStop}%
\bibitem [{\citenamefont
  {ATLAS$\;$Collaboration}(2010{\natexlab{b}})}]{JetEtMiss_Cleaning_Note}%
  \BibitemOpen
  \bibfield  {author} {\bibinfo {author} {\bibnamefont
  {ATLAS$\;$Collaboration}},\ }\href
  {https://atlas.web.cern.ch/Atlas/GROUPS/PHYSICS/RESULTS/summer2010.html}
  {\bibfield  {journal} {\bibinfo  {journal} {ATLAS-CONF-2010-038}\ } (\bibinfo
  {year} {2010}{\natexlab{b}})}\BibitemShut {NoStop}%
\bibitem [{\citenamefont {Sjostrand}\ \emph {et~al.}(2006)\citenamefont
  {Sjostrand}, \citenamefont {Mrenna},\ and\ \citenamefont {Skands}}]{pythia}%
  \BibitemOpen
  \bibfield  {author} {\bibinfo {author} {\bibfnamefont {T.}~\bibnamefont
  {Sjostrand}}, \bibinfo {author} {\bibfnamefont {S.}~\bibnamefont {Mrenna}}, \
  and\ \bibinfo {author} {\bibfnamefont {P.~Z.}\ \bibnamefont {Skands}},\
  }\href@noop {} {\bibfield  {journal} {\bibinfo  {journal} {JHEP}\ }\textbf
  {\bibinfo {volume} {05}},\ \bibinfo {pages} {026} (\bibinfo {year} {2006})},\
  \Eprint {http://arxiv.org/abs/hep-ph/0603175} {arXiv:hep-ph/0603175}
  \BibitemShut {NoStop}%
%%CITATION = HEP-PH/0603175;%%
\bibitem [{\citenamefont
  {ATLAS$\;$Collaboration}(2010{\natexlab{c}})}]{ATLAS_MC09}%
  \BibitemOpen
  \bibfield  {author} {\bibinfo {author} {\bibnamefont
  {ATLAS$\;$Collaboration}},\ }\href@noop {} {\bibfield  {journal} {\bibinfo
  {journal} {ATL-PHYS-PUB-2010-002}\ } (\bibinfo {year}
  {2010}{\natexlab{c}})}\BibitemShut {NoStop}%
\bibitem [{\citenamefont {Sherstnev}\ and\ \citenamefont
  {Thorne}(2008)}]{Sherstnev:2007nd}%
  \BibitemOpen
  \bibfield  {author} {\bibinfo {author} {\bibfnamefont {A.}~\bibnamefont
  {Sherstnev}}\ and\ \bibinfo {author} {\bibfnamefont {R.~S.}\ \bibnamefont
  {Thorne}},\ }\href {\doibase 10.1140/epjc/s10052-008-0610-x} {\bibfield
  {journal} {\bibinfo  {journal} {Eur. Phys. J.}\ }\textbf {\bibinfo {volume}
  {C55}},\ \bibinfo {pages} {553} (\bibinfo {year} {2008})},\ \Eprint
  {http://arxiv.org/abs/0711.2473} {arXiv:0711.2473 [hep-ph]} \BibitemShut
  {NoStop}%
%%CITATION = 0711.2473;%%
\bibitem [{\citenamefont
  {ATLAS$\;$Collaboration}(2010{\natexlab{d}})}]{ATLSIM}%
  \BibitemOpen
  \bibfield  {author} {\bibinfo {author} {\bibnamefont
  {ATLAS$\;$Collaboration}},\ }\href@noop {} {\  (\bibinfo {year} {submitted to
  EPJC, 2010}{\natexlab{d}})},\ \Eprint {http://arxiv.org/abs/1005.4568v1}
  {arXiv:1005.4568v1 [physics.ins-det]} \BibitemShut {NoStop}%
\bibitem [{\citenamefont {Agostinelli}\ \emph {et~al.}(2003)\citenamefont
  {Agostinelli} \emph {et~al.}}]{Agostinelli:2002hh}%
  \BibitemOpen
  \bibfield  {author} {\bibinfo {author} {\bibfnamefont {S.}~\bibnamefont
  {Agostinelli}} \emph {et~al.} (\bibinfo {collaboration} {GEANT4}),\ }\href
  {\doibase 10.1016/S0168-9002(03)01368-8} {\bibfield  {journal} {\bibinfo
  {journal} {Nucl. Instrum. Meth.}\ }\textbf {\bibinfo {volume} {A506}},\
  \bibinfo {pages} {250} (\bibinfo {year} {2003})}\BibitemShut {NoStop}%
%%CITATION = NUIMA,A506,250;%%
\bibitem [{\citenamefont {Catani}\ and\ \citenamefont
  {Seymour}(1997)}]{catani-1998-510}%
  \BibitemOpen
  \bibfield  {author} {\bibinfo {author} {\bibfnamefont {S.}~\bibnamefont
  {Catani}}\ and\ \bibinfo {author} {\bibfnamefont {M.~H.}\ \bibnamefont
  {Seymour}},\ }\href
  {http://www.citebase.org/abstract?id=oai:arXiv.org:hep-ph/9605323} {\bibfield
   {journal} {\bibinfo  {journal} {Nucl. Phys.}\ }\textbf {\bibinfo {volume}
  {B485}},\ \bibinfo {pages} {291} (\bibinfo {year} {1997})}\BibitemShut
  {NoStop}%
\bibitem [{\citenamefont {Nadolsky}\ \emph {et~al.}(2008)\citenamefont
  {Nadolsky} \emph {et~al.}}]{Nadolsky:2008zw}%
  \BibitemOpen
  \bibfield  {author} {\bibinfo {author} {\bibfnamefont {P.~M.}\ \bibnamefont
  {Nadolsky}} \emph {et~al.},\ }\href {\doibase 10.1103/PhysRevD.78.013004}
  {\bibfield  {journal} {\bibinfo  {journal} {Phys. Rev.}\ }\textbf {\bibinfo
  {volume} {D78}},\ \bibinfo {pages} {013004} (\bibinfo {year} {2008})},\
  \Eprint {http://arxiv.org/abs/0802.0007} {arXiv:0802.0007 [hep-ph]}
  \BibitemShut {NoStop}%
%%CITATION = 0802.0007;%%
\bibitem [{\citenamefont
  {ATLAS$\;$Collaboration}(2010{\natexlab{e}})}]{jetProdICHEP}%
  \BibitemOpen
  \bibfield  {author} {\bibinfo {author} {\bibnamefont
  {ATLAS$\;$Collaboration}},\ }\href
  {https://atlas.web.cern.ch/Atlas/GROUPS/PHYSICS/RESULTS/summer2010.html}
  {\bibfield  {journal} {\bibinfo  {journal} {CERN-PH-EP-2010-034}\ } (\bibinfo
  {year} {submitted to EPJC, 2010}{\natexlab{e}})}\BibitemShut {NoStop}%
\bibitem [{\citenamefont {Eichten}\ \emph {et~al.}(1984)\citenamefont
  {Eichten}, \citenamefont {Hinchliffe}, \citenamefont {Lane},\ and\
  \citenamefont {Quigg}}]{Eichten:1984eu}%
  \BibitemOpen
  \bibfield  {author} {\bibinfo {author} {\bibfnamefont {E.}~\bibnamefont
  {Eichten}}, \bibinfo {author} {\bibfnamefont {I.}~\bibnamefont {Hinchliffe}},
  \bibinfo {author} {\bibfnamefont {K.~D.}\ \bibnamefont {Lane}}, \ and\
  \bibinfo {author} {\bibfnamefont {C.}~\bibnamefont {Quigg}},\ }\href
  {\doibase 10.1103/RevModPhys.56.579} {\bibfield  {journal} {\bibinfo
  {journal} {Rev. Mod. Phys.}\ }\textbf {\bibinfo {volume} {56}},\ \bibinfo
  {pages} {579} (\bibinfo {year} {1984})}\BibitemShut {NoStop}%
%%CITATION = RMPHA,56,579;%%
\bibitem [{\citenamefont {Eichten}\ \emph {et~al.}(1986)\citenamefont
  {Eichten}, \citenamefont {Hinchliffe}, \citenamefont {Lane},\ and\
  \citenamefont {Quigg}}]{Eichten:1995akc}%
  \BibitemOpen
  \bibfield  {author} {\bibinfo {author} {\bibfnamefont {E.}~\bibnamefont
  {Eichten}}, \bibinfo {author} {\bibfnamefont {I.}~\bibnamefont {Hinchliffe}},
  \bibinfo {author} {\bibfnamefont {K.~D.}\ \bibnamefont {Lane}}, \ and\
  \bibinfo {author} {\bibfnamefont {C.}~\bibnamefont {Quigg}},\ }\href
  {\doibase 10.1103/RevModPhys.58.1065} {\bibfield  {journal} {\bibinfo
  {journal} {Rev. Mod. Phys.}\ }\textbf {\bibinfo {volume} {58}},\ \bibinfo
  {pages} {1065} (\bibinfo {year} {1986})}\BibitemShut {NoStop}%
%%CITATION = RMPHA,58,1065;%%
\bibitem [{\citenamefont {Chiappetta}\ and\ \citenamefont
  {Perrottet}(1991)}]{Chiappetta1991}%
  \BibitemOpen
  \bibfield  {author} {\bibinfo {author} {\bibfnamefont {P.}~\bibnamefont
  {Chiappetta}}\ and\ \bibinfo {author} {\bibfnamefont {M.}~\bibnamefont
  {Perrottet}},\ }\href@noop {} {\bibfield  {journal} {\bibinfo  {journal}
  {Phys. Lett. B}\ }\textbf {\bibinfo {volume} {235}},\ \bibinfo {pages} {489}
  (\bibinfo {year} {1991})}\BibitemShut {NoStop}%
\bibitem [{\citenamefont {ATLAS$\;$Collaboration}(2010{\natexlab{f}})}]{JES2}%
  \BibitemOpen
  \bibfield  {author} {\bibinfo {author} {\bibnamefont
  {ATLAS$\;$Collaboration}},\ }\href
  {https://atlas.web.cern.ch/Atlas/GROUPS/PHYSICS/RESULTS/summer2010.html}
  {\bibfield  {journal} {\bibinfo  {journal} {ATLAS-CONF-2010-056}\ } (\bibinfo
  {year} {2010}{\natexlab{f}})}\BibitemShut {NoStop}%
\bibitem [{\citenamefont {Neyman}(1937)}]{Neyman}%
  \BibitemOpen
  \bibfield  {author} {\bibinfo {author} {\bibfnamefont {J.}~\bibnamefont
  {Neyman}},\ }\href@noop {} {\bibfield  {journal} {\bibinfo  {journal}
  {Philosophical Transactions of the Royal Society of London A}\ }\textbf
  {\bibinfo {volume} {236}},\ \bibinfo {pages} {333} (\bibinfo {year}
  {1937})}\BibitemShut {NoStop}%
\end{thebibliography}%

% ATLAS authorlist communicated to Andreas by Klaus/Markus
%MS\newpage
%% ATLAS Collaboration author list for 10-SEP-2010
%% Data extracted on 10-Sep-2010 for paperid 37
%% \documentclass[11pt]{article}
%% \usepackage{a4wide}\begin{document}
\begin{flushleft}

\medskip

{\bf The ATLAS Collaboration}

\medskip

G.~Aad$^{\rm 48}$,
B.~Abbott$^{\rm 111}$,
J.~Abdallah$^{\rm 11}$,
A.A.~Abdelalim$^{\rm 49}$,
A.~Abdesselam$^{\rm 118}$,
O.~Abdinov$^{\rm 10}$,
B.~Abi$^{\rm 112}$,
M.~Abolins$^{\rm 88}$,
H.~Abramowicz$^{\rm 153}$,
H.~Abreu$^{\rm 115}$
% ...
\bigskip

% ...
$^{10}$ Institute of Physics, Azerbaijan Academy of Sciences, H. Javid Avenue 33, AZ 143 Baku, Azerbaijan\\
$^{11}$ Institut de F\'isica d'Altes Energies, IFAE, Edifici Cn, Universitat Aut\`onoma  de Barcelona,  ES - 08193 Bellaterra (Barcelona), Spain\\
% ...
$^{48}$ Albert-Ludwigs-Universit\"{a}t, Fakult\"{a}t f\"{u}r Mathematik und Physik, Hermann-Herder Str. 3, D - 79104 Freiburg i.Br., Germany\\
$^{49}$ Universit\'e de Gen\`eve, Section de Physique, 24 rue Ernest Ansermet, CH - 1211 Geneve 4, Switzerland\\
% ...
$^{88}$ Michigan State University, Department of Physics and Astronomy, High Energy Physics Group, East Lansing, MI 48824-2320, United States of America\\
% ...
$^{111}$ University of Oklahoma, Homer L. Dodge Department of Physics and Astronomy, 440 West Brooks, Room 100, Norman, OK 73019-0225, United States of America\\
$^{112}$ Oklahoma State University, Department of Physics, 145 Physical Sciences Building, Stillwater, OK 74078-3072, United States of America\\
% ...
$^{115}$ LAL, Univ. Paris-Sud, IN2P3/CNRS, Orsay, France\\
% ...
$^{118}$ Oxford University, Department of Physics, Denys Wilkinson Building, Keble Road, Oxford OX1 3RH, United Kingdom\\
% ...
$^{153}$ Tel Aviv University, Raymond and Beverly Sackler School of Physics and Astronomy, Ramat Aviv, IL - Tel Aviv 69978, Israel\\
% ...
\end{flushleft}

\end{document}